# Spatial Fuzzy Clustering on Synthetic Aperture Radar Images to Detect Changes

Necmettin BAYAR
Department of
*Telecommunication and electronics Engineering*
Istanbul Technical University
Istanbul, Turkey
bayarn20@itu.edu.tr

W.T AL-SHAIBANI
Department of
*Electrical and Electronics Engineering*
Istanbul Technical University
Istanbul, Turkey
tawfiqwaleed@gmail.com

Ibraheem SHAYEA
Department of
*Telecommunication and electronics Engineering*
Istanbul Technical University
Istanbul, Turkey
ibr.shayea@gmail.com

Abdulkader TAHA
Department of
*Electrical and Electronics Engineering*
kütahya Dumlupıar University
kütahya, Turkey
a.kadereng@yahoo.com

Azızul AZIZAN
Advanced Informatics Department, Razak Faculty of Technology and Informatics
Universiti Teknologi
Kuala Lumpur, Malaysia
azizulazizan@utm.my

*Abstract*— Data and data sources have become increasingly essential in recent decades. Scientists and researchers require more data to deploy AI approaches as the field continues to improve. In recent years, the rapid technological advancements have had a significant impact on human existence. One major field for collecting data is satellite technology. With the fast development of various satellite sensor equipment, synthetic aperture radar (SAR) images have become an important source of data for a variety of research subjects, including environmental studies, urban studies, coastal extraction, water sources, etc. Change detection and coastline detection are both achieved using SAR pictures. However, speckle noise is a major problem in SAR imaging. Several solutions have been offered to address this issue. One solution is to expose SAR images to spatial fuzzy clustering. Another solution is to separate speech. This study utilises the spatial function to overcome speckle noise and cluster the SAR images with the highest achieved accuracy. The spatial function is proposed in this work since the likelihood of data falling into one cluster is what this function is all about. When the spatial function is employed to cluster data in fuzzy logic, the clustering outcomes improve. The proposed clustering technique is used on SAR images with speckle noise to recover altered pixels.

I. INTRODUCTION (HEADING 1)

Change detection is a technique used in remote sensing to identify changes in the same location on the Earth's surface at various periods. Since they are independent from atmospheric and lighting conditions, synthetic aperture radar (SAR) pictures have significant scientific value. Environmental monitoring [1, 2], urban research [3] [4], disaster assessment [5], forest resource monitoring [6] and other study paths have never been easier or more accessible thanks to the growing availability of SAR pictures. However, the background information of SAR images is generally complicated due to the microwave imaging process, and the region's characteristics are relatively jumbled. Structure sensitivity, imaging geometric distortion, imaging system interference, speckle noise and other issues are some challenges. The most challenging issue in the realm of single polarisation SAR image change detection is overcoming speckle noise. The basic unit's phase angle of the SAR imaging system loses continuity due to stochastic backscattering [7], [8]. This results in granular signal intensity-related distortion in pictures. Speckle noise is the intensity distortion associated with the granular signal. The addition of speckle noise to the original picture in the form of multiplication has a significant impact on the interpretation of SAR images. Successfully reducing speckle noise in the field of SAR picture change detection remains to be a difficult challenge.

Clustering is the process of separating data of a desired part without any prior knowledge. Due to enhancements in the artificial intelligence field in recent years, clustering has become an important method for unsupervised learning processes and the clustering of large datasets. Clustering is basically separating data into desired clusters by calculating the probability of each term for each cluster. It is different from classification since classification has a threshold to classify data. Several studies have used unsupervised methods for classifying [9]. Logically, two possibilities are present: high or low. Many clustering algorithms work similarly. They find the exact cluster for every single data. Some apply fuzzy logic where all data consist of clusters by ratio. That means data can have one cluster with α percent and another cluster with β percent, so more clusters can be present. Fuzzy logic will calculate the ratio probability for all clusters, which will allow more flexibility. When separating data to clusters, flexible clustering will provide better results than solid clustering.

Flexible clustering is also used in the Gaussian Mixture Model clustering algorithm. This technique assumes that there exists a finite number of Gaussian distributions, each of which represents a cluster. As a result, data points belonging to a single distribution tend to be grouped together in the Gaussian Mixture Model. For multi-dimensional datasets, being flexible is more accurate than being solid. Clustering algorithm calculates the distance to cluster centres. To minimise the distance between data and cluster centres, the algorithms focus on locating nearby cluster centres. Some algorithms also apply probability throughout this process. For instance, the Expectation Maximisation Algorithm works with the gaussian mixture model to fix data to the model. It calculates the right position of all data on the mixture model, then determines the expectation.

To maximise expectation, it changes the model and position of the data. This paper proposes the spatial function which is the probability of data consisting of one cluster. When the spatial function is applied to data clusters in fuzzy logic, clustering results become better. The proposed clustering algorithm is applied on SAR images to extract the altered pixels that have speckle noise within the images.

The rest of this article is organised as follows: Section II introduces the methodology, Section III provides the dataset and testing procedures, and Section IV presents the conclusion.

## II. METHODOLOGY

This section offers a quick overview of the relevant functions and algorithms utilised in the clustering process. This part provides the foundation knowledge to comprehend the milestones accomplished in this field to date.

### A. Spatial Function

The spatial function determines the likelihood of data belonging to a certain cluster centre. Equation 1 is used to obtain the result [10]:

$$h_{i,j} = \sum_{k \in NB(x_j)} u_{ik} \quad (1)$$

where $NB(x_j)$ denotes the neighbouring members of $x_j$, and $h_{ij}$ signifies the chance that $x_j$ is part of the I cluster centre. In the equation, k is classified as adjacent to $x_j$. That means to calculate the possibility of each data, membership values that have the same intensity with selected pixels are summated. However, instead of adding all membership values of every pixel with the same intensity as the selected pixel, only the neighbouring pixels are summated. If the selected pixel is not in any image edge, the neighbouring pixels will nearly have the same intensity as the selected pixel. The spatial function is then added to the fuzzy clustering algorithm.

### B. Specifications of Spatial Fuzzy Clustering Algorithm

The spatial fuzzy clustering algorithm is proposed to solve the problem of speckle noise [11]. As previously mentioned, the biggest challenge of SAR images is speckle noise. Equations for spatial fuzzy clustering are as follows:

$$u_{i,j} = \frac{1}{\sum_{k=1}^{c} \frac{|x_j - v_i|}{|x_j - v_k|}^{2/(m-1)}} \quad (2)$$

where $i = 1,..., c$ and $j = 1,..., N$. In Equation 2, the membership function is given. It is used to first calculate the membership function in the algorithm. After calculating the spatial function, the new membership function is calculated in Equation 3. M is the weighting exponent of each fuzzy membership:

$$u'_{ik} = \frac{u_{ij}^p h_{ij}^q}{\sum_{k=1}^{c} u_{kj}^p h_{kj}^q} \quad (3)$$

where p and q are the initial fuzzy membership degree and relative weight of the spatial function, respectively. In Equation 3, the membership function is given with the applied spatial function. New centres are calculated with new membership function values. The P and q parameters are the initial fuzzy membership degree and relative weight of the spatial function, respectively.

Once the calculation is complete, new and old membership values are compared to determine the error rate. If the error rate is higher than the defined value, the algorithm recalculates the membership functions until the desired error rate is reached.

Fig. 1 presents the process of the designed spatial fuzzy clustering algorithm. Some parameters can be changed in the algorithm. In future works, optimum parameters will be chosen to analyse the dataset.

### C. Parameter Selection

As mentioned, the spatial fuzzy clustering algorithm possesses several changeable parameters. It depends on the dataset needed to adjust them. Firstly, the p and q parameters are changed. The changed effect on the clustered data is then plotted. The clustered data decreases with increasing p/q ratio.

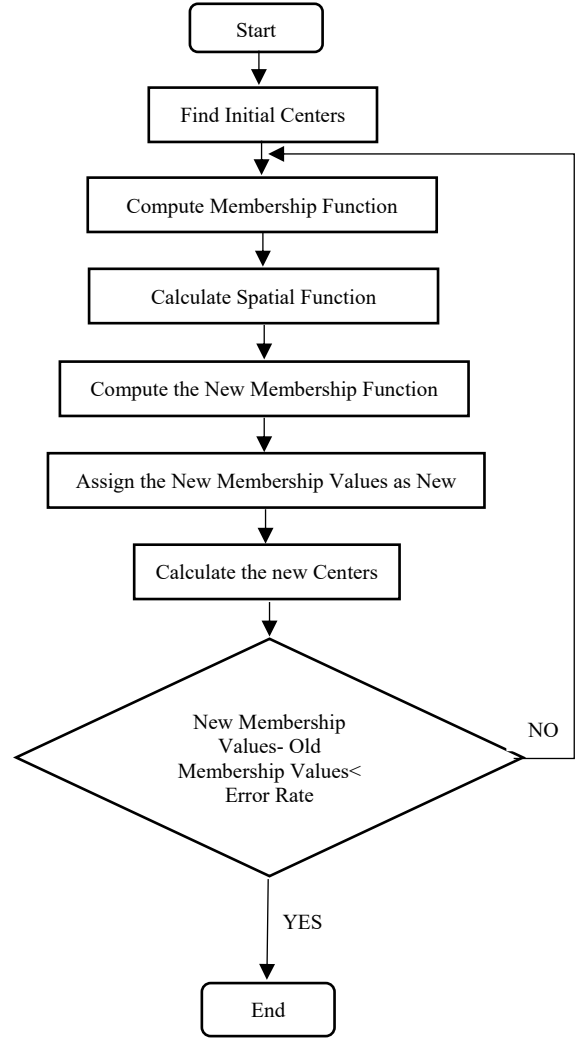

Fig. 1. Spatial Fuzzy Clustering Algorithm

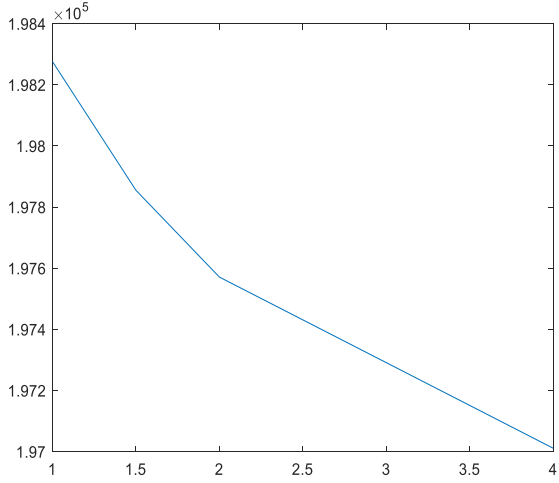

Fig. 2. The P/Q ratio versus the clustered data count

As shown in Fig. 2, the p/q ratio ranges from 1 to 4 and the clustering counts have been plotted accordingly. From the experimental results, it can be seen that the p/q ratio that is closer to one exhibits the best clustering performance.

Another parameter, m, defined in Equation 2 can also be adjusted. To determine the best value for the dataset, several changes have been applied on the m value. The clustering counts have been checked for the single SAR data, as previously accomplished with the p/q ratio. The experimental results revealed that the m parameter value that is higher than 1 relatively provides the same results; therefore, the chosen m parameter is 2.

### D. Generating the Difference Image

The purpose of this paper is to cluster the SAR images. To perform this process, two SAR images of the same location, but taken at different dates, are compared and clustered. In order to compare these images, a difference image is needed [12]. The difference image conceptually shows the difference between two images. There are several ways to generate it. Previous works have used substituting, ratio and logarithmic operations for generating the difference image. This work applies Equation 4, as follows:

$$S_{i,j} = \frac{|I^1_{i,j} - I^2_{i,j}|}{I^1_{i,j} + I^2_{i,j}} \quad (4)$$

where $S_{ij} \in [0, 1]$, and $ij$ represents the grey value of the positions (i, j) in the images at moment t. After applying Equation 4, the grey level image is obtained. The initial cluster centre is required to perform the clustering. To calculate the initial centres, several functions have been used. The experimental results indicate that random initial centres may cause wrong clustering performance, therefore, additional functions should be used to accurately calculate the initial cluster centres.

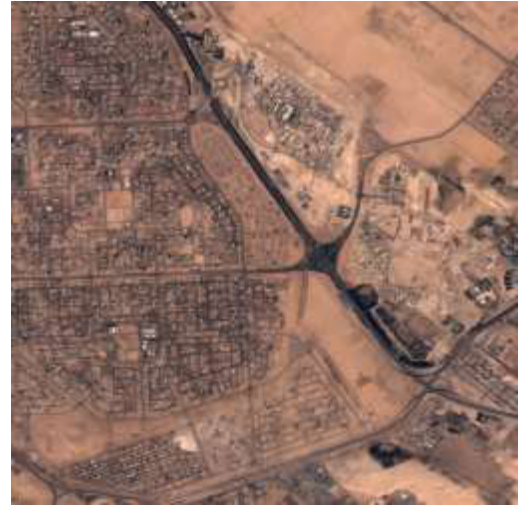

(a)

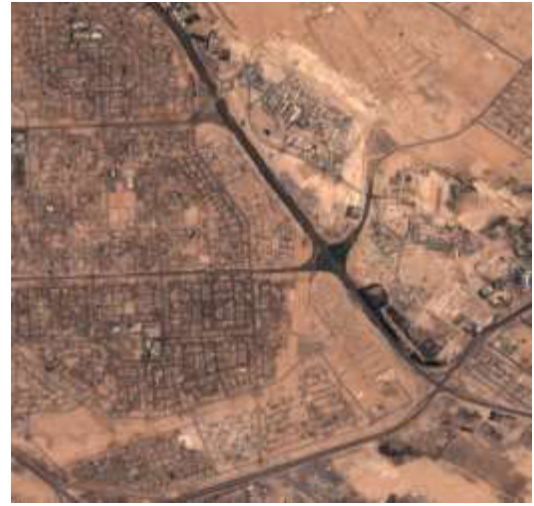

(b)

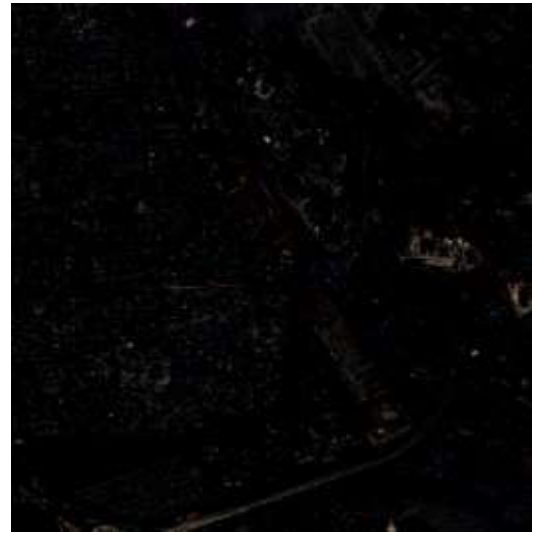

(c)

Fig. 3. Generating the difference image of Abu Dhabi: (a) date: 2016/01/20, (b) date: 2018/03/28 and (c) the generated difference image

## III. DATASET

The Onera Satellite dataset is utilised in this work to cluster the altered pixels in the images. Normally, the proposed technique is limited to SAR pictures. The SAR image datasets for change detection applications are not free. Multispectral pictures are utilised to test the network. Satellites can also take multispectral pictures, therefore, satellite pictures from various dates are included in the collection. They consist of 24 pairs of multispectral pictures collected between 2015 and 2018 by the Sentinel-2 satellites. Locations are chosen from all over the world, including Brazil, the United States, Europe, the Middle East and Asia. The Sentinel-2 satellites delivered registered pairs of 13-band multispectral satellite pictures for each location. Images vary in spatial resolution, ranging between 10 m, 20 m and 60 m.

As previously stated, the Onera Satellite dataset has been utilised to evaluate the clustering method presented in this work. The developed algorithm is fed with the produced difference images from various locations. The clustering performance accuracy cannot be numerically measured since the change detection dataset, which also contains ground truth altering labels, is not free. The pictures that produce different image samples have been used to demonstrate the clustering method's performance. As previously mentioned, the dataset does not contain altered pixels for the entire image. Only changes are labelled in the dataset, which are in urban areas. Thus, clustering accuracy cannot be determined.

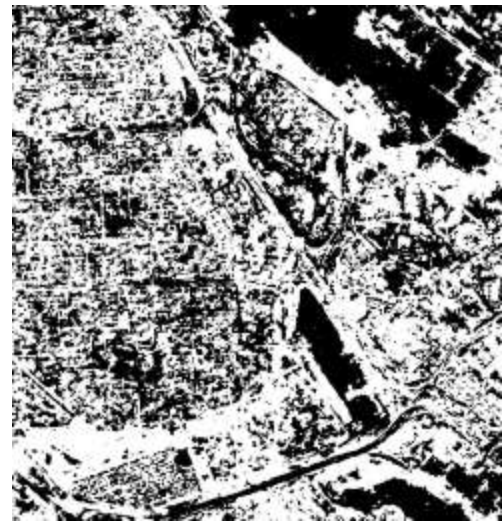

(c)

Fig. 4. Example of clustering results of given images: (a) old image, (b) new image and (c) clustered changed pixels

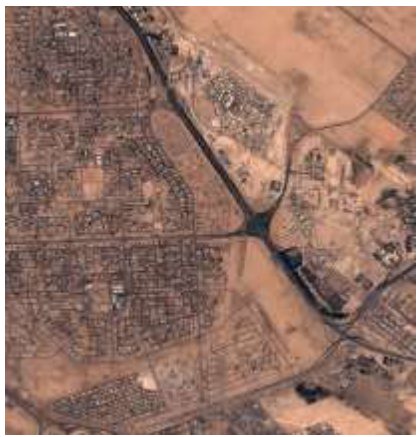

(a)

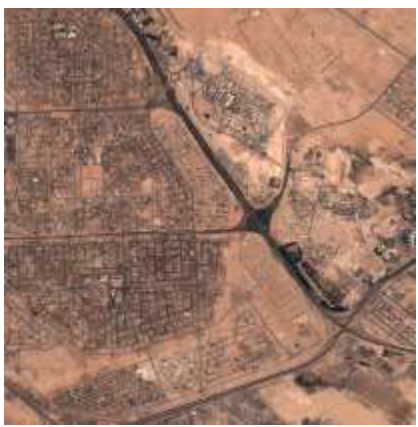

(b)

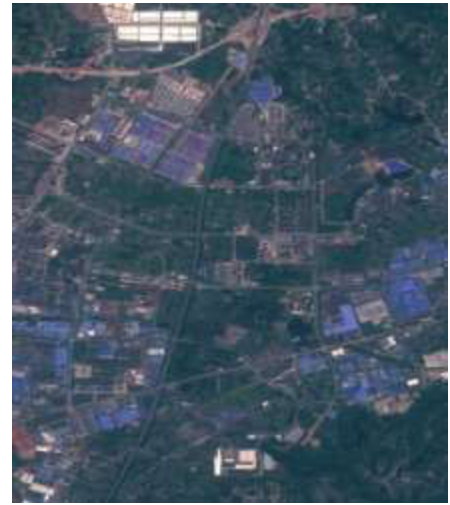

(a)

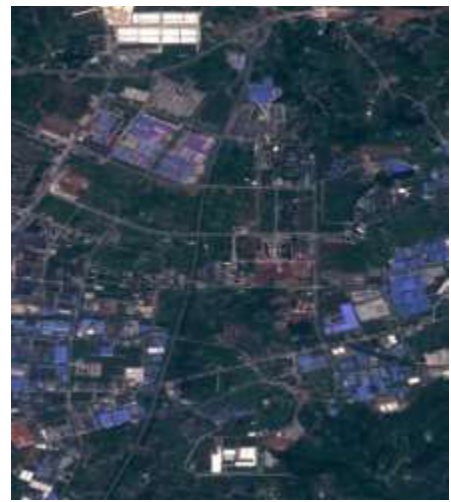

(b)

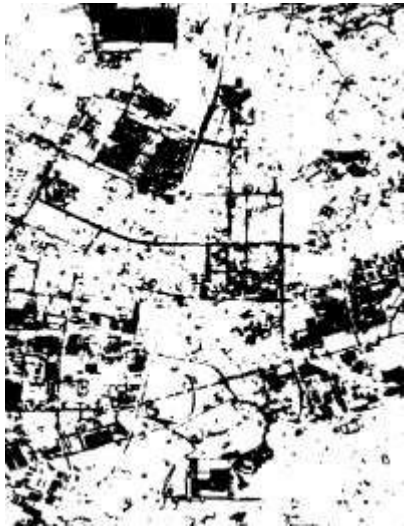

(c)

Fig. 5. Clustering results of given image of Chongqing, China: (a) old image, (b) new image and (c) clustered changed pixels

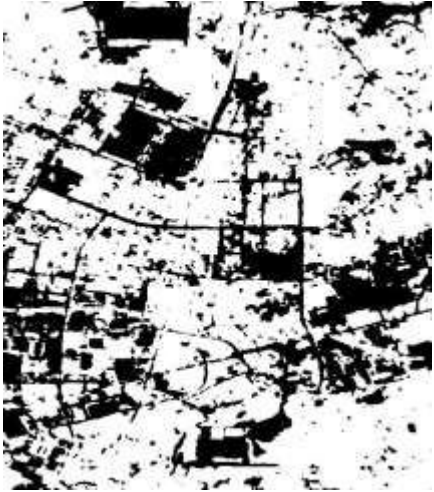

(a)

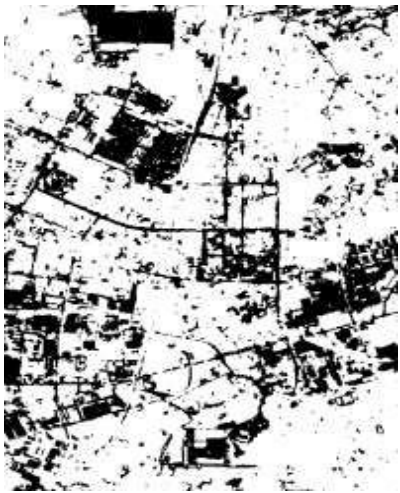

(b)

Fig. 6. Comparison of different types of spatial function calculations: (a) intensity-based spatial function calculation and (b) neighbour pixels based spatial function calculation

## IV. CONCLUSIONS

To overcome speckle noise in pictures, this study used the spatial function to cluster data in the fuzzy clustering method. This approach is applicable for a variety of image types which have speckle noise, including SAR photos, hyperspectral images, multispectral images, etc. The photos acquired from satellites possess significant noise. Previously, spatial functions were computed by adding the membership functions of pixels of the same intensity. The membership function values of neighbouring pixels are summed in this article. The parameters of the spatial fuzzy clustering method were also swept for certain scale in order to identify the optimal values. The Onera Satellite change detection picture dataset was used to test the suggested technique. The outcomes have shown that the proposed approach is more sensitive in small regions.

## ACKNOWLEDGMENT

The authors appreciate the reviewers' valuable time in evaluating this work.